\documentclass[aps,prl,twocolumn,superscriptaddress,amsmath,amssymb]{revtex4-2}
\usepackage{newtxmath}
\usepackage{mathtools}
\DeclarePairedDelimiter\abs{\lvert}{\rvert}
\DeclarePairedDelimiter\norm{\lVert}{\rVert}
\DeclareMathOperator{\Tr}{Tr}
\usepackage[colorlinks,linkcolor=magenta,citecolor=blue]{hyperref}

\newcommand{\ket}[1]{\left\lvert\smash{#1}\right\rangle}

\newcommand{\kket}[1]{\left\lVert\smash{#1}\right\rangle}

\newcommand{\op}[2]{\vphantom{#1#2}\left\lvert\smash{#1}\middle\rangle\!\middle\langle\smash{#2}\right\rvert}

\newcommand{\braket}[2]{\left\langle\smash{#1}\middle\vert\smash{#2}\right\rangle}

\newcommand{\mel}[3]{\vphantom{#1#2#3}\left\langle\smash{#1}\middle\vert\smash{#2}\middle\vert\smash{#3}\right\rangle}

\newcommand{\dd}[1]{\mathrm{d}#1}
\newcommand{\dv}[2]{\frac{\mathrm{d}#1}{\mathrm{d}#2}}

\newcommand{\Lcal}{\mathcal{L}}
\newcommand{\Hhat}{\hat{H}}
\newcommand{\hhat}{\hat{h}}
\newcommand{\Lhat}{\hat{L}}
\newcommand{\rhoh}{\hat{\rho}}
\newcommand{\sigmah}{\hat{\sigma}}
\newcommand{\ini}{\mathrm{ini}}
\newcommand{\ness}{\mathrm{NESS}}
\newcommand{\gbs}{\mathrm{Gibbs}}
\newcommand{\sectionprl}[1]{{\par\textit{#1}.---}}
\begin{document}

\title{Tensor-network approach to thermalization in open quantum many-body systems}

\author{Hayate Nakano}
\email{hayate.nakano@phys.s.u-tokyo.ac.jp}
\affiliation{Department of Physics, University of Tokyo, Tokyo 113-0033, Japan}

\author{Tatsuhiko Shirai}
\email{tatsuhiko.shirai@aoni.waseda.jp}
\affiliation{Department of Computer Science and Communications Engineering, Waseda University, Tokyo 162-0042, Japan}

\author{Takashi Mori}
\email{takashi.mori.fh@riken.jp}
\affiliation{RIKEN Center for Emergent Matter Science (CEMS), Wako 351-0198, Japan}

\date{\today}

\begin{abstract}
We investigate the relaxation dynamics of open non-integrable quantum many-body systems in the thermodynamic limit by using a tensor-network formalism. 
We simulate the Lindblad quantum master equation (LQME) of infinite systems by making use of the uniform matrix product operators (MPO) as the ansatz of their density matrices.
Furthermore, we establish a method to measure the thermodynamic equivalence between two states described by the uniform MPOs.
We numerically show that when an initial state of the LQME is a thermal Gibbs state, a time evolved state is always indistinguishable from a Gibbs state with a time-dependent effective temperature in the weak-dissipation and thermodynamic limit.
\end{abstract}

\maketitle

\sectionprl{Introduction}
Thermalization of \textit{isolated} quantum many-body systems has attracted much interest~\cite{Rigol2008-tk,Gogolin2016-wf,Mori2018-kv,Abanin2019-rr}, which was triggered by experimental realizations of well-controlled and well-isolated quantum systems in ultracold atoms~\cite{Trotzky2012-mi,Kaufman2016-uw,Neill2016-bq,Hubig2015-zq}.
From a theoretical side, important concepts such as typicality and eigenstate thermalization hypothesis (ETH) have been established to explain this phenomenon~\cite{Neumann1929-dj,Deutsch1991-nj,Srednicki1994-kr}.

Recently, it has been shown that these concepts are also useful in understanding non-equilibrium dynamics of \textit{open} quantum many-body systems under dissipation via non-equilibrium environments or under continuous quantum measurements~\cite{Shirai2020-zz,Ashida2018-mf,Lange2018-vd}.
Under suitable conditions, dissipative quantum dynamics is described by the Lindblad quantum master equation (LQME)~\cite{Lindblad1976-lv}.
However, because of the lack of the detailed balance condition, there is no simple yet general description of the non-equilibrium steady state (NESS) and relaxation dynamics towards it.
It has been recently argued that thermal Gibbs states emerge in the course of relaxation dynamics when the Hamiltonian of the system of interest obeys the ETH and the dissipation is weak enough~\cite{Shirai2020-zz,Ashida2018-mf}.

These results in the previous works are based on exact numerical calculations for relatively small systems.
However, there is a subtlety here related to the exchangeability of the two limits, i.e., the thermodynamic limit and the weak-dissipation limit.
The emergence of thermal Gibbs states are easily derived if we take the weak-dissipation limit before the thermodynamic limit.
However, in considering a thermodynamically large system, we should take the thermodynamic limit first and then take the weak-dissipation limit.
The emergence of thermal Gibbs states is highly nontrivial in the latter limiting procedure.
Since it is difficult to distinguish the two limiting procedures in numerical calculations of finite systems, it is desired to directly calculate the non-equilibrium dynamics in the thermodynamic limit to establish the emergence of thermal Gibbs states in non-equilibrium open quantum many-body systems.

In this letter, we do it by utilizing the tensor-network (TN) formalization~\cite{Orus2014-xl,Bridgeman2017-xw,Orus2019-bb}.
There are several TN-based algorithms which can simulate infinite systems directly, e.g., infinite density matrix renormalization group (iDMRG)~\cite{McCulloch2008-lw,Schollwock2011-cw}, infinite time-evolving block decimation (iTEBD)~\cite{Vidal2007-fv}, infinite projected entangled pair state (iPEPS)~\cite{Jordan2008-or}, time-dependent variational principle (TDVP), and variational uniform matrix product state algorithm (VUMPS)~\cite{Haegeman2011-jb,Zauner-Stauber2018-bm,Vanderstraeten2019-fo}.
In these algorithms, we use a TN to represent a quantum state vector, not a density operator.
By regarding operators as states via the Choi--Jamio\l{}kowski isomorphism~\cite{Choi1975-po,Jamiolkowski1972-gl}, we can apply these algorithms to simulate quantum many-body operators~\cite{Zwolak2004-wy,Mascarenhas2015-bj,Cui2015-ps}.
In this letter, we use the uniform matrix product operator (MPO)~\cite{Verstraete2004-fd} as an ansatz to represent a density matrix of a one-dimensional translation-invariant quantum many-body system and directly simulate the LQME in the thermodynamic limit.

By making use of the method mentioned above, we show numerical evidences of the following claim:
When an initial state at time $t=0$ is given by a thermal Gibbs state, the state at time $t>0$ is indistinguishable from a Gibbs state with a time-dependent effective temperature in the weak-dissipation limit after the thermodynamic limit.
Furthermore, we will show that the effective temperature dynamics is governed by a simple ordinary differential equation, which is efficiently solvable by a TN-based approach.

\sectionprl{LQME and time-dependent effective temperature}
We assume that a density matrix $\rhoh$ of a macroscopic system in contact with an environment is described by the LQME defined as follows:
\begin{align}
\dv{\rhoh}{t} &= \Lcal_{\gamma}[\rhoh] = \Lcal_{(0)}[\rhoh] + \gamma \Lcal_{(1)}[\rhoh]\label{eq:defLQME}\\
&\coloneqq -i[\Hhat,\rhoh] + \gamma \sum_i\left( \hat{L}_i\rhoh\hat{L}_i^\dagger -\frac{1}{2} \{ \hat{L}_i^\dagger \hat{L}_i, \rhoh \}\right).\nonumber
\end{align}
Here, $\Hhat$ denotes the Hamiltonian of the system.
Operators $\{\Lhat_i\}$ are called Lindblad operators, which characterize the dissipation due to the interaction between the system and the environment.
We assume $\Hhat$ is a sum of translation-invariant neargerst-neighbor interactions
\begin{align}
    \Hhat &= \sum_i \hhat_{i,i+1},\label{eq:defH}
\end{align}
and the Lindblad operator $\Lhat_i$ acts only on the site $i$. 
We write the time evolution of the LQME as $e^{\Lcal_\gamma t}[\rhoh^\ini]$.

Informally, our claim is expressed as follows:
When the initial state $\rhoh^\ini$ is a thermal Gibbs state at an effective temperature $(\beta^\ini)^{-1}$,
the subsequent state $e^{\Lcal_\gamma t}[\rhoh^\ini]$ is equivalent to a time-dependent Gibbs state $\rhoh^\gbs_{\beta_\gamma(t)} \coloneqq e^{-\beta_{\gamma}(t) \Hhat}/\Tr[e^{-\beta_\gamma(t)\Hhat}]$ with a time-dependent effective temperature $\beta_\gamma(t)$ in the thermodynamic limit followed by the weak-dissipation limit ($\lim_{\gamma\rightarrow0}\lim_{V\rightarrow\infty}$).
Here, we say two states are ``equivalent'' if they are indistinguishable by any measurement of local operators.
As a sufficient condition of this indistinguishability, we consider vanishinig of the R\'enyi-2 divergece density~\cite{Mori2016-xk}:
\begin{align}
    s_2(\sigmah \Vert \rhoh) \coloneqq f(\sigmah^2 \rhoh^{-1}),\label{eq:defs2}
\end{align}
where $f(\rhoh)$ is defined as
\begin{align}
    f(\rhoh) \coloneqq \lim_{V\rightarrow\infty}\frac{1}{V}\ln \Tr[\rhoh].\label{eq:deff}
\end{align}
In particular, we denote the R\'enyi-2 divergence with respect to a thermal state $\hat{\rho}_\beta^{\mathrm{Gibbs}}$ by
\begin{align}
    s^\gbs_2(\sigmah;\beta) &\coloneqq s_2(\sigmah \Vert \rhoh^{\gbs}_\beta).\label{eq:defs2beta}
\end{align}

Now, we can write down the precise expression of our claim:
\begin{align}
    &s_2^\gbs(\rhoh^\ini ; \beta^\ini) = 0\nonumber \\
    &\Rightarrow \lim_{\gamma\to +0}\sup_{t>0}s^\gbs_2\left(e^{\Lcal_\gamma t}[\rhoh^{\ini}]; \beta_\gamma(t)\right) = 0,\label{eq:mainclaim}
\end{align}
where $\beta_\gamma$ evolves as
\begin{align}
    \dv{\beta}{t} =  - \gamma \frac{c(\beta)}{\chi(\beta)},\quad \beta(0)=\beta^{\mathrm{ini}}.\label{eq:betaclaim}
\end{align}
Here, we define $c(\beta)$ as the rate of change of energy density per $\gamma$ and $\chi(\beta)$ as the specific heat of the system as
\begin{align}
    &c(\beta) \coloneqq \frac{1}{V\gamma} \Tr\left[\Hhat\dv{\rhoh^\gbs_\beta}{t}\right] = \frac{1}{V}\Tr[\Hhat\Lcal_{(1)}[\rhoh^\gbs_\beta]]\label{eq:defc}\\
    &=\frac{1}{V} \sum_k \Tr\left<\left[\hat{L}_k^\dagger,\Hhat\right]\hat{L}_k\right>_\beta= \sum_{k\in\{0,1\}}\left<\left[\hat{L}_k^\dagger,\hhat_{0,1}\right]\hat{L}_k \right>_\beta \nonumber,\\
    &\chi(\beta) \coloneqq \frac{1}{V} \left(\left< \Hhat^2 \right>_{\beta} - \left< \Hhat \right>_{\beta}^2\right)\label{eq:defchi}\\
    &= \sum_k \left(\left< \hhat_{0,1} \hhat_{k,k+1} \right>_\beta - \left<\hhat_{0,1}\right>_\beta \left< \hhat_{k,k+1} \right>_\beta\right)\nonumber,
\end{align}
where $\left<\bullet\right>_\beta$ denotes the thermal average $\Tr[\rhoh^\gbs_\beta \bullet]$. We have used the translation invariance in the last equality.

By applying separation of variables, we can obtain an integral equation equivalent to Eq.~\eqref{eq:betaclaim} as
\begin{align}
    \int_{\beta^{\ini}}^{\beta_\gamma(t)} \dd{x} \frac{\chi(x)}{-c(x)} = \gamma t\label{eq:betaint}.
\end{align}
Therefore, we can calculate $\beta_\gamma(t)$ by evaluating the left hand side integral numerically.
For general $\Hhat$, we can prove that $c(\beta)=0$ has at least one positive solution by showing $c(0) \leq 0$ and $c(+\infty) \geq 0$ (see Appendix C in \cite{Shirai2020-zz}).
We define $\beta^\ness$ as the solution of $c(\beta)=0$ since $\beta^\ness = \beta_\gamma(+\infty)$ holds.

\sectionprl{Tensor-network for infinite open system}
In this letter, we use TN-based algorithms to test our hypotheses numerically.
In recent years, several TN formalisms for open quantum many-body systems has been proposed~\cite{Jaschke2018-iu,Weimer2019-nk}.
Among them, we adopt the matrix product operator (MPO) representation for our calculation~\cite{Zwolak2004-wy,Cui2015-ps,Gangat2017-jf}.
MPO is a standard TN representation of quantum many-body operators in one-dimensional lattice systems.
One pro of MPO is that we can directly represent states in the thermodynamic limit under translation invariance.
Besides MPO, there are other powerful TN representations called matrix product density operator (MPDO)~\cite{Verstraete2004-fd} and locally purified tensor-network (LPTN)~\cite{Werner2016-wp}.
MPDO and LPTN are better than MPO in some perspectives since positivity and hermicity of the density matrix are guaranteed.
However, they can exhibit low expressive power compared with MPO.
For instance, when a density matrix $\rhoh$ is given by MPO with bond-dimension $D$, the bond-dimension of MPDO required to approximate $\rhoh$ cannot be bounded by $D$~\cite{De_las_Cuevas2013-af}. 

\begin{figure}[tbp]
\centering
\includegraphics[width=\linewidth]{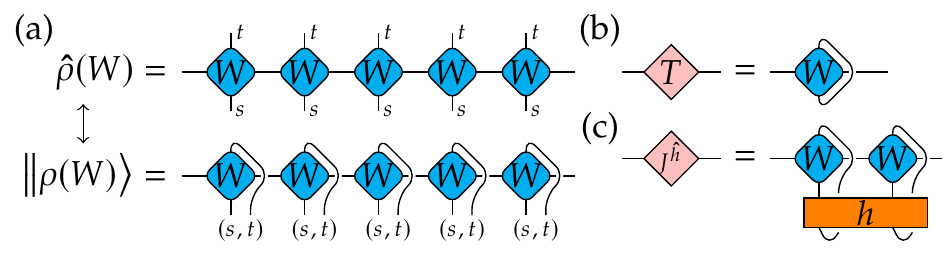}
\caption{
(a) Schematic picture of Choi--Jamio\l{}kowski isomorphism between MPO and MPS (b) (c) Diagramatic representation of transfer matrix and operator transfer matrix
\label{fig:tndiag}}
\end{figure}

For a translation-invariant system, we define a uniform MPO state
\begin{align}
    &\rhoh(W) = \sum_{\mathclap{\{s\},\{t\}}} \Tr\left[\prod_{i=1}^V W^{s_i,t_i}\right] \op{\{s\}}{\{t\}}\label{eq:defMPO}\\
    &\eqqcolon \sum_{\mathclap{\{s\},\{t\},\{a\}}} W^{s_1,t_1}_{a_1,a_2} W^{s_2,t_2}_{a_2,a_3} \ldots W^{s_V,t_V}_{a_V,a_1} \op{s_1,\ldots,s_V}{t_1,\ldots,t_V}\nonumber
\end{align}
with a rank 4 tensor $W \in \mathbb{C}^{D\times D\times d\times d}$, where $D$ denotes the bond-dimension and $d$ denotes the dimension of the local Hilbert space.
For a given MPO state, we define a transfer matrix and its spectral decomposition as follows:
\begin{align}
    T = \sum_{s} W^{s,s} = \sum_{i=1}^D \lambda_i |i)(i|,\quad \abs{\lambda_1} > \abs{\lambda_2} \geq \ldots,\label{eq:defT}
\end{align}
where $(i|$ and $|i)$ denote left and right eigenvector of $T$ satisfying $(i|j) = \delta_{ij}$.
We note that $|i)^\dagger$ is not equal to $(i|$ in general since $T$ can be a non-Hermitian matrix.
$\lambda_1$, the eigenvalue with the largest magnitude, corresponds to the function $f(\rhoh)$ defined in Eq.~\eqref{eq:deff} as
\begin{align}
    f(\rhoh) = \ln \abs{\lambda_1}\label{eq:feqln}.
\end{align}
Hereafter, we always assume $\lambda_1=1$ for normalized density matrices.

For a local operator $\hat{o}$ acting on sites between $i$ and $i+N$, we define the operator transfer matrix as
\begin{align}
    J^{\hat{o}} \coloneqq \sum_{\{s\},\{t\}}\left(\prod_{j=i}^{i+N} W^{s_j,t_j}\right) \mel{t_i,\dots,t_{i+N}}{\hat{o}}{s_i,\dots,s_{i+N}}\label{eq:defJ}.
\end{align}
Diagramatic representations of these transfer matrices are illustrated in Fig.~\ref{fig:tndiag} (b), (c).
Then, the expectation value of $\hat{o}$ can be written as
\begin{align}
    \left<\hat{o}\right> = \lim_{V\rightarrow\infty}\Tr[\hat{\rho}(W)\hat{o}] = (1|J^{\hat{o}}|1)\label{eq:valofop}.
\end{align}
We can also calculate the specific heat $\chi$, which contains non-local terms, by using $J^{\hhat}$ as
\begin{align}
    \chi &= \left< \left\{ \hhat_{-1,0}, \hhat_{0,1} \right\} \right>+\left<\hhat_{0,1}^2\right>-3\left<\hhat_{0,1}\right>^2\nonumber\\
    &\hphantom{=} + 2(1| J^{\hhat} (1 - T)^+ J^{\hhat} |1), \label{eq:valofchi}
\end{align}
where the symbol $\bullet^+$ denotes the Moore--Penrose generalized matrix inverse of $\bullet$. It is noted that the following equation holds:
\begin{align}
    (1 - T)^+ \coloneqq \sum_{i=2}^{D} \frac{1}{1-\lambda_i} |i)(i| = \sum_{k=0}^\infty \left(T^k - |1)(1|\right).\label{eq:defpinv}
\end{align}

We can easily obtain an MPO of the product of two density matrices represented by MPOs.
When two density matrices are given as $\rhoh(W_1)$ and $\rhoh(W_2)$,
an MPO of $\rhoh(W_1)\rhoh(W_2)$ is given by a tensor $(W_{3})^{s,t}_{(a_1,b_1),(a_2,b_2)} = \sum_u (W_1)_{a_1,a_2}^{s,u} (W_2)_{b_1,b_2}^{u,t}$.
Thus, we can calculate $s_2(\rhoh \Vert \sigmah)$ by MPOs of $\rhoh^{-1}$ and $\sigmah$.
In general, it is a difficult task to calculate an MPO of $\rhoh(W)^{-1}$ by $W$, but fortunately, for the Gibbs state, we can use the equation $(\rhoh^\gbs_\beta)^{-1} = \rhoh^\gbs_{-\beta}\Tr[e^{\beta\Hhat}]\Tr[e^{-\beta\Hhat}]$ to obtain the MPO of the inverse.
By substituting it into Eq.~\eqref{eq:defs2}, we can deform the definition of the divergence density as: 
\begin{align}
    s^\gbs_2(\sigmah;\beta) = f(\sigmah^2\rhoh^\gbs_{-\beta}) + f(e^{-\beta\Hhat}) + f(e^{\beta\Hhat}).\label{eq:valofs2}
\end{align}

\sectionprl{Algorithms for MPO}
Now, we have to simulate time evolutions of density matrices represented by MPO.
For this purpose, we consider the mapping between the MPO \eqref{eq:defMPO} and the matrix product state (MPS) of the system with ancilla sites by the Choi--Jamio\l{}kowski isomorphism ($\rhoh = \sum_{i}p_i\op{\psi_i}{\psi_i}\rightarrow \kket{\rho} = \sum_{i} p_i \ket{\psi_i}\otimes\ket{\psi_i}$) as showin in Fig.~\ref{fig:tndiag} (a):
\begin{align}
    \rhoh(W) \rightarrow \kket{\rho(W)} = \sum_{\{(s,t)\}} \Tr\left[\prod_{i=1}^V W^{s_i,t_i} \right] \ket{\{s\}}\otimes\ket{\{t\}}.\label{eq:choi}
\end{align}
Then, we can apply many kinds of algorithms to simulate real/imaginary time evolutions of MPS.

In this letter, we use a time-dependent variational principle (TDVP) algorithm~\cite{Haegeman2011-jb,Vanderstraeten2019-fo}.
In the TDVP, we solve the optimization problem of $\Delta W$ to minimize $\norm{\kket{\rho(W+\Delta W)} - e^{\Delta t \Lcal_\gamma}\kket{\rho(W)}}^2$ for a time-step $\Delta t$.
By repeating this step, we can project the LQME dynamics onto the subspace represented by the uniform MPS.
We also use the TDVP algorithm to obtain Gibbs states by considering the imaginary-time evolution of $\Hhat$.

In these simulations, we have to take care of the normalization.
As mentioned earlier, the normalization of $\rhoh(W)$, i.e., $\Tr[\rhoh(W)]=1$, is determined by the eigenvalues of the transfer matrix \eqref{eq:defT}.
On the other hand, the TDVP algorithm assumes that the state vector is normalized in the sence of the inner-product norm, i.e., $\braket{\rho(W)}{\rho(W)}=\Tr[\rhoh(W)\rhoh(W)^\dagger]=1$.
We use the state vector normalization in the simulation and convert it to the density matrix normalization for calculating physical quantities.
In the simulation of the imaginary time evolution, the normalization breaks after each time step.
We re-normalize the vectors for each time and memory the ratios.
As can be seen in Eq.~\eqref{eq:valofs2}, we need to calculate the trace of $e^{\pm \beta \Hhat}$.
We can obtain it by multiplying the ratios.

\sectionprl{Numerical Results}
Let us consider a quantum Ising model with a longitudinal magnetic field
\begin{align}
    \Hhat = \sum_i g \hat{Z}_{i} \hat{Z}_{i+1} + \Delta \hat{Z}_i + \Omega \hat{X}_i,\quad \hat{L}_i = \hat{X}_i - i \hat{Y}_i,\label{eq:defqising}
\end{align}
where $\hat{X}$, $\hat{Y}$ and $\hat{Z}$ denotes the Pauli matrices.
In this letter, we fix the parameters as $(g,\Delta,\Omega) =(1,0.9045,0.8090)$.
In this parameter set, this model satisfies the strong ETH~\cite{Kim2014-ad}.
This system can be realized in Rydberg atom systems~\cite{Carr2013-kk,Letscher2017-hs}, and novel features like non-equilibrium phase transitions under dissipation have been discussed~\cite{Lee2011-vy}.

\begin{figure}[tbp]
\centering
\includegraphics[width=\linewidth]{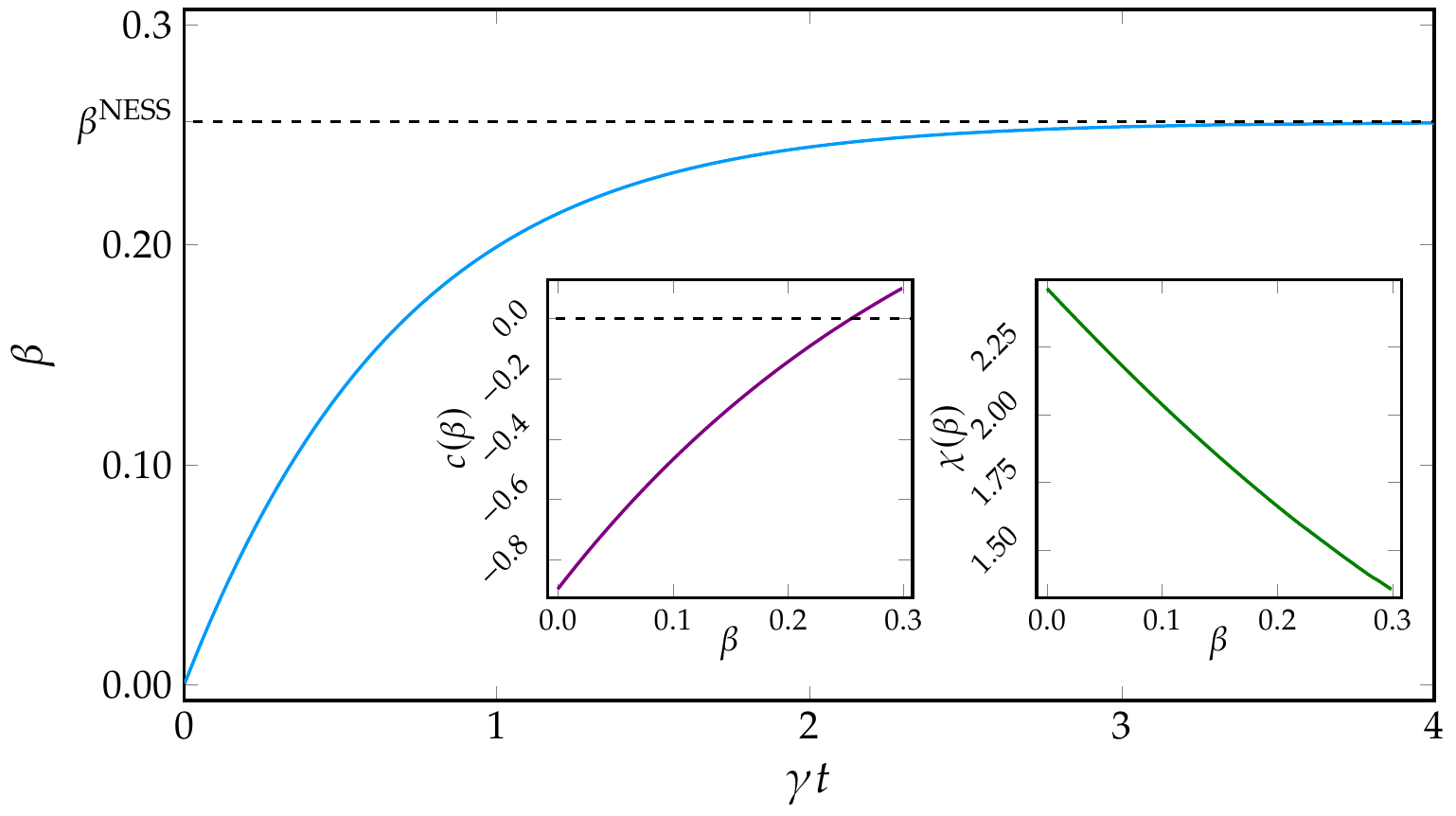}
\caption{
Main: plot of $\beta(t)$ obtained by a numerical integration of Eq.~\eqref{eq:betaint}. The horizontal dashed line denotes $\beta^\ness$ obtained by solving $c(\beta)=0$.
Inset: plots of $c(\beta)$ and $\chi(\beta)$ defined in Eqs.~\eqref{eq:defc} and \eqref{eq:defchi}. For each $\beta$, $c(\beta)$ and $\chi(\beta)$ was calculated by the operator transfer matrix formula defined in Eqs.~\eqref{eq:valofop} and \eqref{eq:valofchi}.
\label{fig:betagammat}
}
\end{figure}

In Fig.~\ref{fig:betagammat}, we show $\beta_\gamma(t)$ obtained by numerical integration of $\chi(\beta)/c(\beta)$,
where $c(\beta)$ and $\chi(\beta)$ obtained by the Gibbs states approximated by MPO.
$\beta^\ness$, the solution of $c(\beta) = 0$, is around $0.256$.
Since we focus on a relatively high-temperature regime,
we do not have to take the bond-dimension so large~\cite{Hastings2006-ki,Kliesch2014-jf,Molnar2015-su,Kuwahara2020-sd}.
Indeed, we obtained almost equivalent results by taking $D=4$ and $D=8$.
The accuracy of the result is dominated by not the bond-dimension but the imaginary time-step $\Delta \beta$.
In our study, we checked that $\Delta \beta = 10^{-7}$ is sufficiently small.

\begin{figure}[tbp]
\centering
\includegraphics[width=\linewidth]{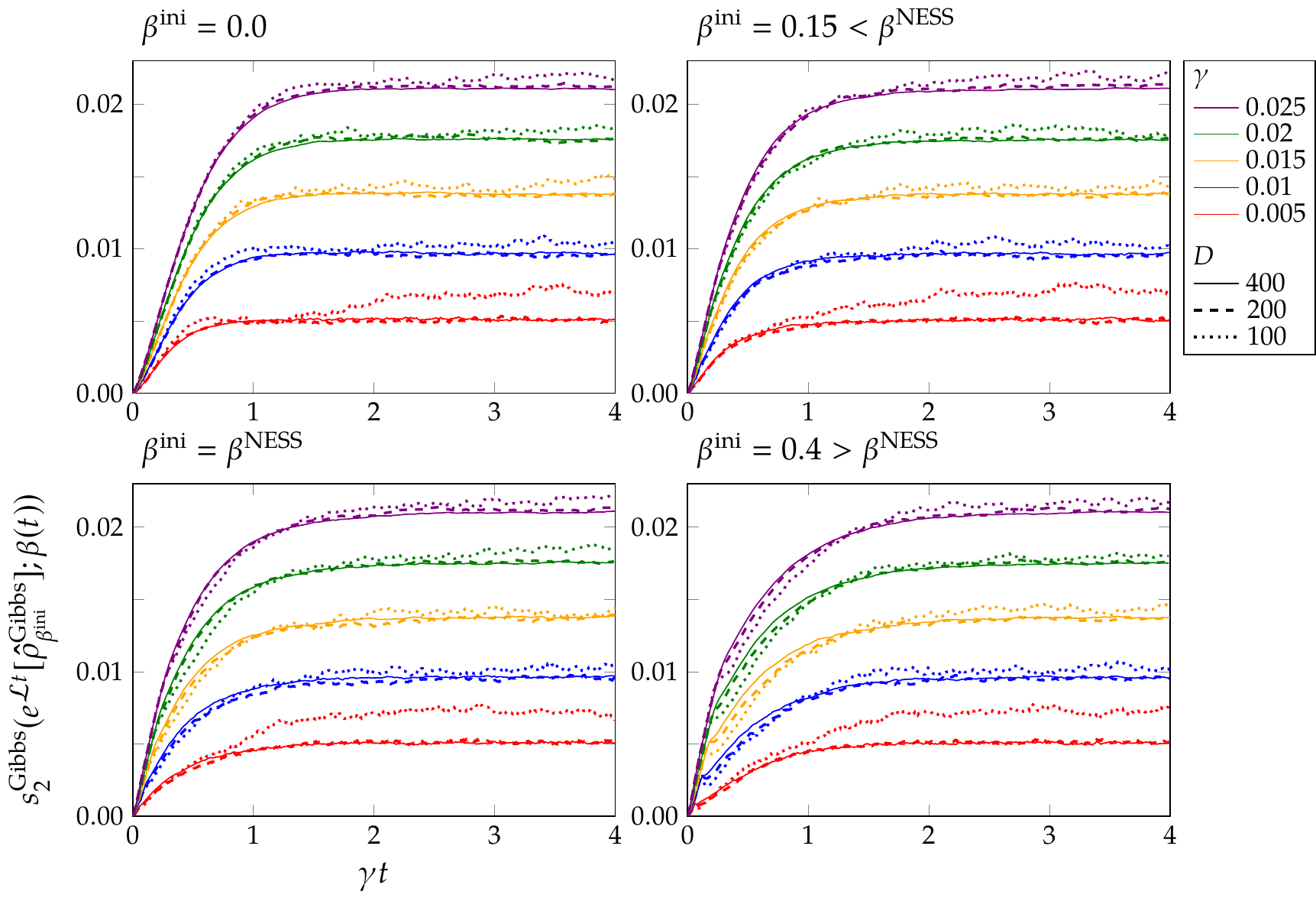}
\caption{
Time evolutions of R\'enyi-2 divergece density $s_2^\gbs$ defined in \eqref{eq:defs2beta} for several initial temperatures 
simulated by the TDVP algorithm with time-step $\Delta t=0.2$. $D$ denotes the bond-dimension of MPO.
\label{fig:gamma_s2}}
\end{figure}

\begin{figure}[tbp]
\centering
\includegraphics[width=\linewidth]{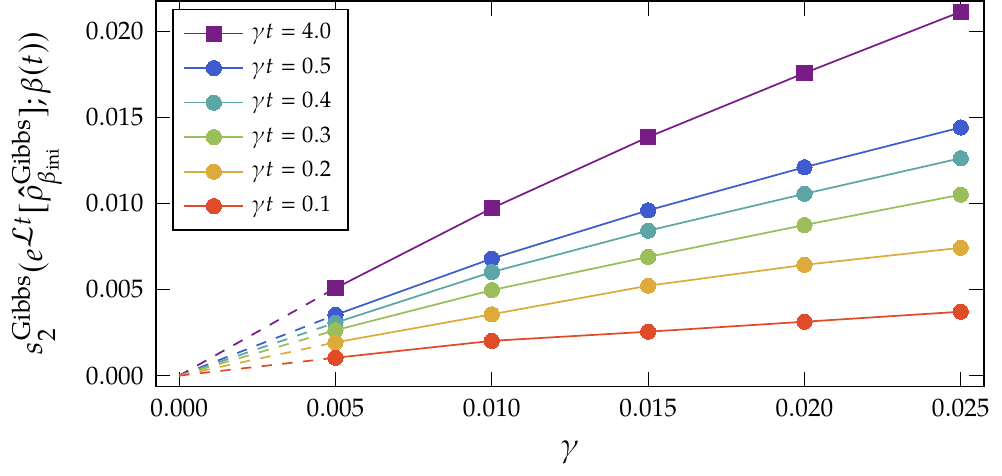}
\caption{
$\gamma$ dependence of R\'enyi-2 divergece density $s_2^\gbs$ with $\beta^\ini = \beta^\ness$ for several fixed $\gamma t$.
Lines with circles correspond to the states in the middle of relaxation, and lines with squares correspond to the NESS. Dasshed lines connect points on $\gamma=0.005$ and the origin ($\gamma=0,s_2=0$) with a straight line.\label{fig:fixt}}
\end{figure}

We show a numerical evidence of our hypothesis in Eq.~\eqref{eq:mainclaim} in Figs.~\ref{fig:gamma_s2}, \ref{fig:fixt}.
We take $\rhoh^\ini = \rhoh^\gbs_{\beta^\ini}$ for $\beta^\ini = 0.0,0.15,\beta^\ness(\approx 0.256),0.4$ and $\gamma=0.005,0.01,0.015,0.02,0.025$ to calculate $s_2^\gbs$ defined in Eq.~\eqref{eq:defs2beta}.
There are two parameters we have to choose for numerical calculation:
the maximum bond-dimension $D$ and the time-step $\Delta t$.
We sweep the maximum bond-dimension $D = 100,200,400$ and show all the results in Fig.~\ref{fig:gamma_s2}.
We can see the results of $D=200,400$ meet well and we can say the result converges well for $D=400$.
We also sweep the time-step $\Delta t$ and found $\Delta t=0.2$ is small enough.
We show $\gamma$ dependences of $s_2^\gbs$ for fixed $\gamma t$ in Fig.~\ref{fig:fixt}.
We can see the linear dependence on $\gamma$ in small $\gamma$ region ($\gamma < 0.01$) for each $\gamma t$.

Remarkably, we succeeded in simulating the whole relaxation dynamics to the NESS.
In many cases, TN-based simulations of real-time dynamics of quantum many-body systems succeed only in a short-time regime because of the linear growth of the entanglement entropy during the unitary time evolution.
In LQME, the unitary time evolution is described by the term $\Lcal_{(0)}$ defined in Eq.~\eqref{eq:defLQME}.
In our case, since we take the initial state as the Gibbs state, which satisfies $\Lcal_{(0)}[\rhoh^\gbs] = 0$, the growth of entanglement by unitary time evolution is suppressed.
Dissipation shifts the state gradually away from the Gibbs state,
but it also causes relaxation of matrix elements of the density matrix to a single steady state.
It is assumed that the balance between these two effects of the dissipation enables us to simulate the relaxation dynamics based on the TN, unlike the thermalization of isolated quantum many-body systems from a quenched initial state.

\sectionprl{Discussion}
First, we point out that the series of our claims in this letter are complementary to previous studies analyzing NESS~\cite{Shirai2020-zz}.
In the previous study, the density matrix of the NESS was obtained by perturbation expansion of static equation $\Lcal_\gamma[\rhoh] = 0$, and strong ETH was used to show that the NESS is indistinguishable from a thermal state.
In contrast, our formulation enables us to discuss not only the NESS but also the overall relaxation dynamics for a thermal Gibbs initial state.

It is an important future problem to understand what happens if the condition of the initial state is relaxed.
In this case, numerical calculations based on TN become challenging since the entanglement growth occurs from the beginning of the time evolution.
Indeed, we simulated a time evolution of a finite-size system starting from a non-thermal state by exact diagonalization and found a peak structure of the entanglement.
This structure can be regarded as a result of the balance between the entanglement growth caused by the unitary time evolution and the suppression caused by the dissipation.
Furthermore, this peak becomes large in the weak-dissipation limit, which we are interested in.
One possible solution is to consider an initial state with a small difference from the thermal Gibbs state and investigate how the peak structure depends on the initial state and the dissipation.

Another important future problem is proving our claim rigorously.
As some previous studies~\cite{Ashida2018-mf,Shirai2020-zz} suggest, ETH can be useful for this purpose.

\begin{acknowledgments}
The authors thank Synge Todo for the fruitful discussion. 
H. N. was supported by Advanced Leading Graduate Course for Photon Science (ALPS), the University of Tokyo.
T. S. and T. M. were supported by Japan Society for the Promotion of Science KAKENHI
Grants No. 18K13466 and No. 19K14622, respectively.
\end{acknowledgments}

%\bibliography{Paperpile}

%begin supp.tex
%\clearpage
%\onecolumngrid
%\setcounter{equation}{0}
%\setcounter{figure}{0}
%\setcounter{table}{0}
%\setcounter{page}{1}
%\renewcommand{\theequation}{S\arabic{equation}}
%\renewcommand{\thefigure}{S\arabic{figure}}
%\renewcommand{\bibnumfmt}[1]{[S#1]}
%\renewcommand{\citenumfont}[1]{S#1}
%\makeatother
%\begin{center}
%  \textbf{\Large Supplemental Materials}
%  \bigskip
%  Hayate Nakano, Tatsuhiko Shirai and Takashi Mori \\
%  \textit{Department of Physics, University of Tokyo, Tokyo 113-0033, Japan}\\
%  \textit{Green Computing Systems Research Organization, Waseda University, Tokyo 162-0042, Japan}\\
%  \textit{RIKEN Center for Emergent Matter Science (CEMS), Wako 351-0198, Japan}
% \end{center}

\end{document}